\documentclass[journal]{IEEEtran}
\makeatletter
\def\ps@headings{%
\def\@oddhead{\mbox{}\scriptsize\rightmark \hfil \thepage}%
\def\@evenhead{\scriptsize\thepage \hfil \leftmark\mbox{}}%
\def\@oddfoot{}%
\def\@evenfoot{}}
\makeatother
\usepackage{graphicx,subfigure}
\usepackage{url}
\usepackage[utf8]{inputenc}
\usepackage[ruled,vlined]{algorithm2e}
\usepackage{algpseudocode}

\usepackage{tikz}
\newcommand\copyrighttext{
	\footnotesize \textcopyright This work has been submitted to the IEEE for possible publication. Copyright may be transferred without notice, after which this version may no longer be accessible.}
\newcommand\copyrightnotice{
	\begin{tikzpicture}[remember picture,overlay]
	\node[anchor=south,yshift=10pt] at (current page.south) {\fbox{\parbox{\dimexpr\textwidth-\fboxsep-\fboxrule\relax}{\copyrighttext}}};
	\end{tikzpicture}
}
\usepackage{listings}
\usepackage{amssymb,amsfonts}
\usepackage[cmex10]{amsmath}
\usepackage{gensymb}
\usepackage{color}
\usepackage{upquote}
\usepackage{algpseudocode}
\usepackage[square, comma, sort&compress, numbers]{natbib}
\usepackage{multirow}
\usepackage{booktabs}
\usepackage{tabu}
\usepackage{colortbl}

\newtheorem{theorem}{\bf Theorem}
\definecolor{intnull}{RGB}{213,229,255}
\definecolor{c01}{RGB}{50,200,50}
\definecolor{c02}{RGB}{100,100,100}
\newcommand{\tOED}{\texttt{OED}}

\newcommand{\tGreedy}{\texttt{Greedy}}
\newcommand{\tILP}{\texttt{ILP}}
\newcommand{\trILP}{\texttt{rILP}}
\newcommand{\cR}{\mathcal{R}}
\newcommand{\cRgreedy}{\mathcal{R}_{greedy}}
\newcommand{\tNULL}{\texttt{NULL}}
\title{Entanglement Distribution in Satellite-based
Dynamic Quantum Networks\thanks{Alena Chang, Yinxin Wan, Guoliang Xue,
and Arunabha Sen are all affiliated with
School of Computing and Augmented Intelligence,
Arizona State University,
Tempe, AZ 85287.
Emails: \{ahchang, ywan28, xue, asen\}@asu.edu.
Alena Chang and Yinxin Wan made
equal contributions to this paper.
This paper was supported in part by NSF grants 2007083 and 2007469.
The information reported herein does not reflect the position
of the policy of the funding agency.
}
}
\author{Alena Chang, {\em Student Member}, {\em IEEE},
Yinxin Wan, {\em Student Member}, {\em IEEE},
Guoliang Xue, {\em Fellow}, {\em IEEE},
Arunabha Sen, {\em Senior Member}, {\em IEEE}}
\begin{document}
\maketitle
\copyrightnotice
\begin{abstract}
Low Earth Orbit (LEO) satellites present a compelling opportunity for the
establishment of a global quantum information network. 
However, satellite-based entanglement distribution from a networking
perspective has not been fully investigated.
Existing works often do not account for satellite movement over time when
distributing entanglement and/or often do not permit entanglement distribution
along inter-satellite links, which are two shortcomings we address in this paper.
We first define a system model which considers both satellite movement over
time and inter-satellite links.
We next formulate the optimal entanglement distribution ($\tOED$) problem
under this system model and show how to convert the $\tOED$ problem in a
dynamic physical network to one in a static logical graph which can be used
to solve the $\tOED$ problem in the dynamic physical network.
We then propose a polynomial time greedy algorithm for computing
satellite-assisted multi-hop entanglement paths.
We also design an integer linear programming (ILP)-based algorithm to compute optimal solutions
as a baseline to study the performance of our greedy algorithm.
We present evaluation results to demonstrate the advantage of our model and
algorithms.
\end{abstract}
%%%%%%%%%%%%%%%%%%%%%%%%%%%%%%%%%%%%%%%%%%%%%%%%%%%%%%
%%%%%%%%%%%%%%%%%%%%%%%%%%%%%%%%%%%%%%%%%%%%%%%%%%%%%%

%===============================================================================
\section{Introduction}
\label{sec:Intro}
%===============================================================================
%
\noindent
Predicated entirely on entanglement, quantum networks enable Alice
to securely send information to Bob by teleporting the state of a qubit from
her site to that of Bob without physically transferring the qubit itself,
consuming the entanglement in the process. 
Entanglement is thus regarded as a precious resource in quantum communications,
a currency dubbed {\em ebits}. 
The book by Van Meter~\cite{van2014quantum} is considered an authoritative
text on quantum networks, while Khatri and Wilde~\cite{https://doi.org/10.48550/arxiv.2011.04672} provide a thorough
mathematical treatment of quantum communications.

Repeaters in a quantum network employ local operations and classical
communication (LOCC) to manipulate the state of one or more qubits in order
to facilitate end-to-end entanglement between two parties who may not be
directly connected by a physical communication link. 
We first generate entanglements between adjacent repeaters
along links to form a repeater chain of entangled photon pairs between Alice
and Bob. 
We then perform entanglement swapping~\cite{9798254}
by applying a Bell state measurement (BSM) at each of the intermediate
repeaters, transforming two consecutive entanglements into one, until
Alice and Bob directly share an entangled photon pair.
Alice and Bob may now use this ebit to perform quantum teleportation.
The entire process is assisted by classical heralding signals which indicate whether an attempt at entanglement generation or entanglement swapping is successful.
% %

Through a strictly terrestrial lens, entanglement distribution on a global
scale will always be hindered by photon loss over optical fibers, whose
entanglement distribution rate decays exponentially with distance and is
upper bounded by the repeaterless bound~\cite{PLOB2017}.
Satellites, particularly Low Earth Orbit (LEO) satellites, offer a promising
workaround, as their low altitude makes entanglement distribution demonstrably
achievable and their ability to travel in a periodic manner reduces the number
of ground repeaters required to generate entanglement between two remote
parties. 
A comprehensive review of the state-of-the-art work in space quantum
communications can be found in~\cite{SatQIN2023, Sidhu2021Advances}.
Entanglement distribution in a space-ground scheme can be accomplished
in different ways, which we describe in the following.

The scheme proposed in~\cite{PhysRevA.91.052325} describes a
{\em double downlink configuration}, in which satellites armed with
entanglement sources beam down entangled photon pairs to two ground
repeater stations at a time. 
Each ground repeater houses a number of quantum non-demolition (QND)
measurement devices and quantum memories (QMs). 
If a satellite successfully transmits an entangled photon pair to two
ground repeaters, with each repeater in possession of one photon, the
successful attempt is heralded by the repeaters' QND devices. 
The repeaters then store the photons in their respective QMs until they
receive information about successful entanglement distribution at
neighboring repeaters, at which point they may perform BSMs to extend
entanglements as needed.

The more recent architecture studied
in~\cite{Gundogan2021Proposal, Liorni2021QuantumRI} allows for both ground
and space repeaters. 
Satellites are not only fitted with entanglement sources, but also QND
devices and QMs, so entanglement swapping can be executed in space as well
as on Earth.
Likewise, in addition to QND devices and QMs, ground stations may also
be equipped with entanglement sources, so ground stations can distribute
entanglement to satellites. 
In addition to {\em downlink channels}, these modifications allow for
{\em uplink channels} and
{\em inter-satellite links}~\cite{Sidhu2021Advances}.

Note that none of the aforementioned configurations require ground stations
to directly transmit ebits to each other, which spares us the burden of
mitigating photon loss over optical fibers.
Our objective is to distribute ebits to a pair of ground stations purely by
way of LEO satellites.
This approach allows us to avoid optical fibers entirely in favor of
free-space optical links in vacuum and only two atmospheric channels. 
The entanglement distribution problem we study is in the
vein of~\cite{pant2019routing, UCSC}, 
but we shift our attention to the celestial so as to align ourselves with
the most probable future direction in quantum
communications~\cite{SatQIN2023, Sidhu2021Advances}.

Existing studies on satellite-based entanglement distribution from a
networking standpoint often do not consider satellite movement over time
and/or do not permit entanglement distribution along inter-satellite links. 
A satellite may be within communication range of a ground station or another
satellite for them to share an entangled photon pair at one point in time,
but they may be too far away from each other at a different point in time. 
A system model studying LEO satellite-based entanglement distribution
should factor in the limitations on feasible photon transmission imposed
by satellite movement.
However, this has not been well investigated from a networking perspective.

LEO satellite-based entanglement distribution which excludes
inter-satellite links poses a major hindrance due to the fact that the
communication range of a single LEO satellite, given its low altitude,
cannot accommodate two ground stations whose distance exceeds the communication range. 
If a ground station in New York City (NYC) and a ground station in
Singapore request an ebit, a single LEO satellite cannot honor their
request.

Prior networking papers on satellite-based entanglement distribution include~\cite{Khatri2021Spooky, 9798300, Wallnofer2022Simulating}. 
The system model defined in~\cite{Khatri2021Spooky} lays fundamental groundwork for modeling satellite-based entanglement distribution, taking into account satellite movement over time. 
However, the model only considers downlinks and does not permit inter-satellite links.
Similarly,~\cite{9798300} only considers the double downlink configuration and also does not account for satellite movement. 
While~\cite{Wallnofer2022Simulating} enables inter-satellite links and considers satellite movement, it focuses on a specific scenario consisting of three satellites connecting two ground stations, in which the central satellite is positioned halfway between the ground stations.

To the best of our knowledge, we are the first to investigate satellite-based entanglement distribution from a networking perspective at scale while taking into account satellite movement over time and inter-satellite links. 
The contributions of this paper are fourfold:
\begin{itemize}
\item
We propose a system model of an LEO satellite-based quantum network which
factors in both satellite movement over time and inter-satellite links,
under which we define the optimal entanglement distribution ($\tOED$) problem.
    
\item
We introduce the concept of logical graphs and demonstrate how to construct
the {\em static} logical graph corresponding to a set of connection requests
in a {\em dynamic} physical network.
    
\item
We design a polynomial time greedy algorithm for solving the $\tOED$ problem.

\item
We conduct extensive performance evaluation to demonstrate the advantage
of our model and algorithms.
\end{itemize}

The remainder of this paper is organized as follows. 
In Section~\ref{sec:02}, we define the system model and the $\tOED$ problem. 
In Section~\ref{sec:03}, we describe the concept of logical graphs as
a means to solve the $\tOED$ problem. 
In Section~\ref{sec:04}, we present our algorithms for the $\tOED$ problem.
In Section~\ref{sec:05}, we present evaluation results together with our
observations and analysis.
Section~\ref{sec:06} concludes the paper.
%%%%%%%%%%%%%%%%%%%%%%%%%%%%%%%%%%%%%%%%%%%%%
%%%%%%%%%%%%%%%%%%%%%%%%%%%%%%%%%%%%%%%%%%%%%

%==============================================================================
\section{System Model}
\label{sec:02}
\noindent
%==============================================================================
In this section, we present the system model for the space-ground integrated
quantum network.
We also formulate the entanglement distribution problem to be studied.

%==============================================================================
\subsection{Ground Stations and Moving Satellites}
\label{sec:02A}
\begin{figure}[htbp!]
    \centering
    \includegraphics[width=3.4in]{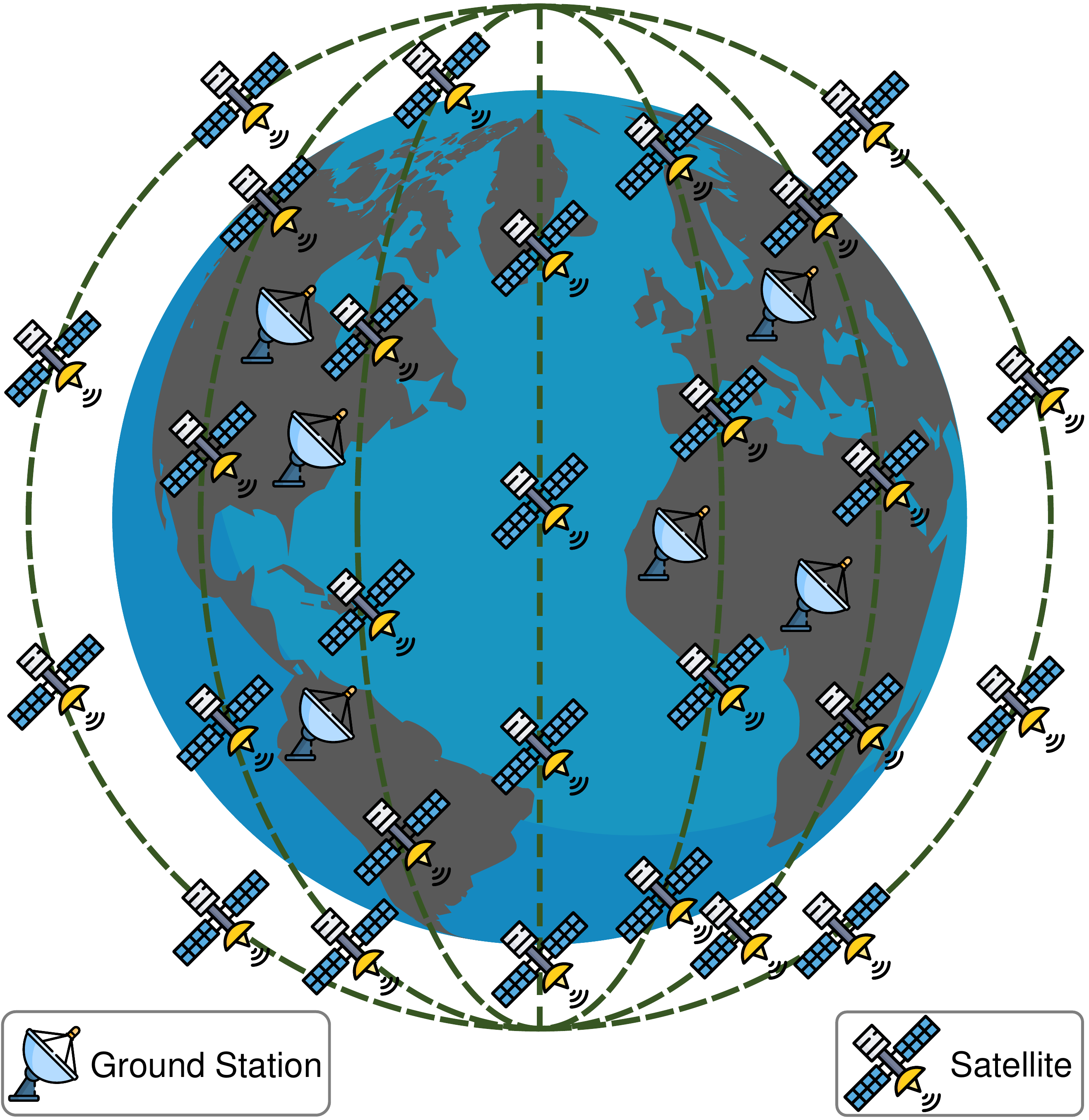}
    \caption{An example of the type of network we study with $R=6$ and $K=8$.}
    \label{fig:sys}
\end{figure}

%==============================================================================
\noindent
There are $R$ equally spaced rings of satellites, and each ring passes over
the North and South Poles in an arrangement known as a polar orbit.
Each ring contains $K$ evenly spaced satellites.
The satellites are arranged so that they do not collide at the Poles. 
This kind of satellite constellation is known as a Walker Star
constellation~\cite{walker1970circular} and has been adopted in space-based entanglement distribution work such as~\cite{Khatri2021Spooky}.
Different from existing models~\cite{9798300, Wallnofer2022Simulating},
we offer an at-scale model taking into consideration the movement of satellites.
Consequently, the geolocation of a satellite dynamically changes and is
determined by time, denoted by $\tau$,
where $\tau=0$ means midnight,
$\tau=1$ means $1$:$00$AM, etc.
In contrast, the geolocation of a ground station is considered stationary.
Fig.~\ref{fig:sys} illustrates an example of the type of network we
study, with $R=6$ and $K=8$.
In the figure, $28$ satellites and $6$ ground stations are visible,
while there are a total of $48$ satellites.

Equipped with transmitters, receivers, and QMs, a satellite and another
satellite or a ground station can establish a link if they are within
each other's communication range.
This distinguishes our model from existing models~\cite{Khatri2021Spooky,9798300}
in that we allow inter-satellite links.

%==============================================================================
\subsection{Problem Formulation}
\label{02B}
%==============================================================================
%
\noindent
A connection request $r_i$ is specified by a $4$-tuple
$(s_i, t_i, d_i, w_i)$, where
$s_i$ is the \emph{source} node which corresponds to a ground station,
$t_i$ is the \emph{destination} node which corresponds to another ground station,
$d_i > 0$ is the \emph{demand} for this connection request which quantifies
the number of ebits to be transmitted,
and $w_i$ is the \emph{reward} for this connection request (if served).
To serve connection request $r_i$, we need to find an $s_i$-$t_i$ path $\pi_i$
of bandwidth $d_i$ via one or more satellites where all the satellites as well
as $s_i$ and $t_i$ have sufficient transmitters, receivers, and QMs to
support such an entanglement path of bandwidth $d_i$.
If such a path does not exist, request $r_i$ cannot be served.
If such a path exists, then the transmitters, receivers, QMs,
and channels corresponding to $\pi_i$ need to be reserved
for $r_i$ and must not be shared with any other connection request.

We assume that connection requests come in and are served in batches. 
Let $\cR = \{r_1, r_2, \ldots, r_N\}$ be a given set of connection requests
and $[\tau, \tau + \delta]$ is the time interval in which the communication
links used in the entanglement paths should remain operational.
Let $\cR'$ be a subset of $\cR$.
We say $\cR'$ is \emph{feasible}, if we can find a path $\pi_i$ to serve $r_i$
for each $r_i \in \cR'$ such that
(1) all of the links used in the paths $\pi$ are operational in the entire time
interval $[\tau, \tau + \delta]$,
and
(2) for any two requests $r_{i} \in \cR'$ and $r_{j} \in \cR'$,
path $\pi_{i}$ and path $\pi_{j}$ do not share
any transmitters/receivers/QMs/channels.
The reward for serving $\cR'$ is
$w(\cR') = \sum_{r_i \in \cR'} w_i$.

The \textbf{Optimal Entanglement Distribution} (\tOED) problem seeks to find
a feasible subset $\cR_{opt}$ of $\cR$ with maximum total reward.
In other words, $\cR_{opt} \subseteq \cR$ is feasible and
$w(\cR_{opt})$ is at least as large as $w(\cR')$ for any $\cR' \subseteq \cR$
that is feasible.
%%%%%%%%%%%%%%%%%%%%%%%%%%%%%%%%%%%%%%%%%%%%%%%%%%%%%
%%%%%%%%%%%%%%%%%%%%%%%%%%%%%%%%%%%%%%%%%%%%%%%%%%%%%

%==============================================================================
\section{The Logical Graph}
\label{sec:03}
%==============================================================================
\noindent
The $\tOED$ problem involves establishing entanglements using links between
two nodes whose distance changes over time (due to the movement of the satellites)
and the links used are guaranteed to be operational in the entire time interval
$[\tau,\! \tau + \delta]$.
Therefore, we are dealing with a network that is \emph{dynamic}, in the
sense that a link between two nodes may exist at one time, and may not
exist at another time.
As a means to solve the $\tOED$ problem in a dynamic network, we introduce
the notion of a logical graph corresponding to a given set of connection requests
$\cR$ in the following.
Therefore we will have a \emph{physical network} that is dynamic, and a
\emph{logical graph} that is static.

Assume that the set of connection requests $\cR$ is given.
We construct a logical graph $G=(V, E)$ for $\cR$ as follows.
For each node $z$ (a satellite or a ground station) in the physical network,
there is a corresponding vertex $l(z)$ in the logical graph.
We use the notation $l(z)$ to mean that
\emph{node} $z$ in the physical network
corresponds to
\emph{vertex} $l(z)$ in the logical graph.
For each vertex $v$ in the logical graph, we use $p(v)$ to denote the
corresponding node in the physical network.
In other words, $l(\cdot)$ is a one-to-one mapping from the set of nodes
in the physical network to the set of vertices in the logical graph;
$p(\cdot)$ is a one-to-one mapping from the set of vertices in the logical
graph to the set of nodes in the physical network.

For two vertices $u$ and $v$ in $V$, there is an undirected edge $(u, v) \in E$
if and only if $p(u)$ and $p(v)$ are within their communication range in the
entire time interval $[\tau, \tau + \delta]$.
For any given time $\tau$, we know the coordinates of $p(u)$ and the coordinates
of $p(v)$.
There is an edge $(u, v) \in E$ if and only if
the \emph{maximum distance between $p(u)$ and $p(v)$ in the interval $[\tau, \tau + \delta]$} is within the communication range of $p(u)$ and $p(v)$.

If $(u, v) \in E$ is an edge, we know that $p(u)$ and $p(v)$ are within their
communication range in the entire interval $[\tau, \tau + \delta]$.
The \emph{bandwidth} of edge $(u, v)$ is the number of available channels
between $p(u)$ and $p(v)$.
The number of transmitters/receivers/QMs at vertex $v \in V$ is
the same as that in $p(v)$.

%==============================================================================
\begin{figure}[htbp!]
    \centering
    \includegraphics[width=3.4in]{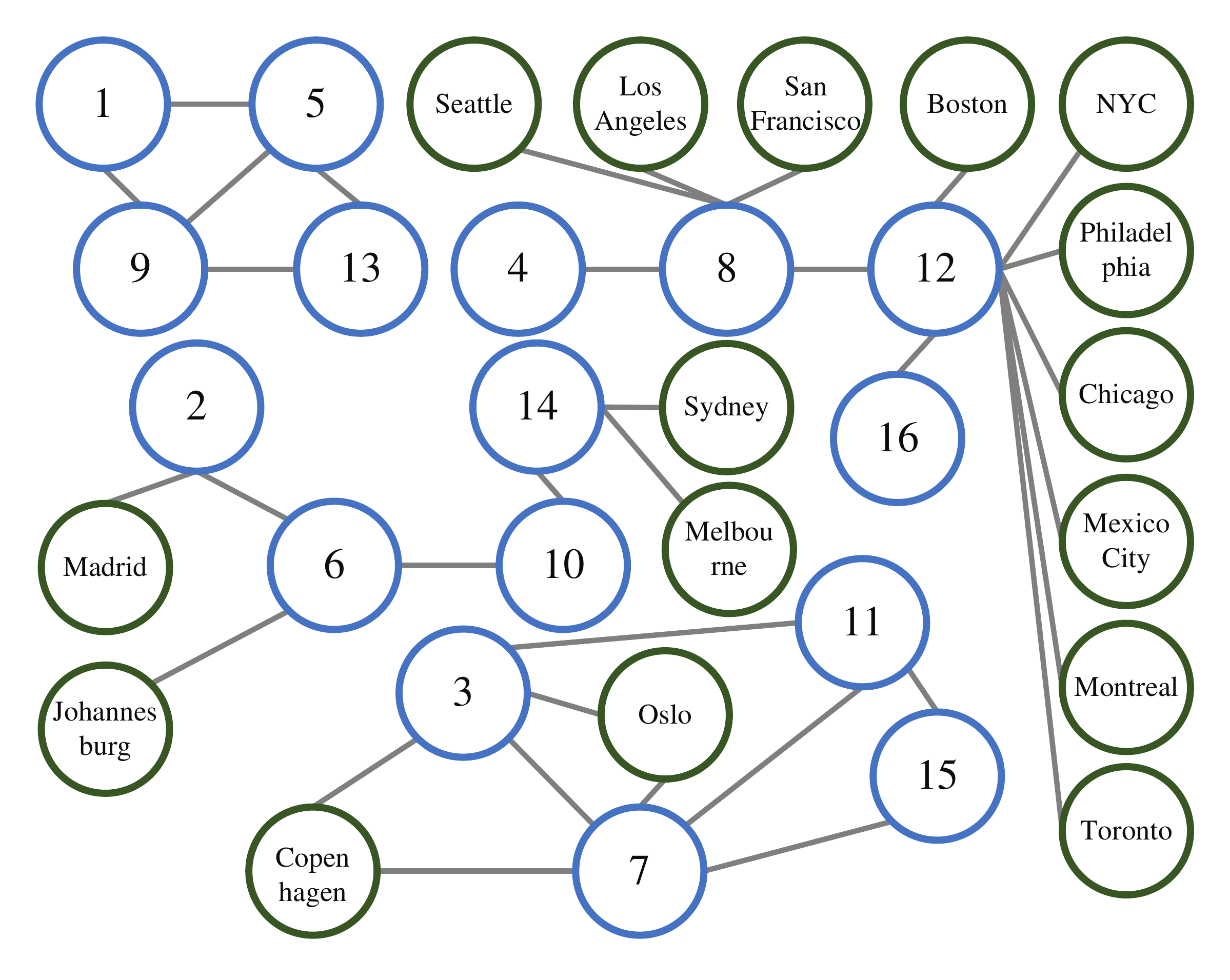}
    \caption{An example of a logical graph where $R=K=4$.}
    \label{fig:logical}
\end{figure}
%==============================================================================

Fig.~\ref{fig:logical} illustrates a part of a logical graph computed
from the physical network where $R=K=4,~\tau=1$, and $\delta=0.01$.
There are a total of $60$ ground stations in the physical network, all
residing in major cities across the world.
In the computed logical graph, $31$ ground stations have edges connecting
them to satellites.
To enhance readability, we only show $16$ ground stations
while omitting $15$ (all in Europe, and connected to satellites $S_3$ and $S_7$).
There are $16$ edges in the logical graph where
both constituent vertices correspond to satellites in the physical network.
These edges correspond to inter-satellite links in the physical network.
We observe that it is possible to establish an entanglement between the
ground station in Madrid and the ground station in Sydney using the path
Madrid-$S_2$-$S_6$-$S_{10}$-$S_{14}$-Sydney.
Such a connection is impossible without using inter-satellite links.

From the definition of the logical graph, we have the following theorem
whose proof is straightforward and hence omitted.
\begin{theorem}
\label{the:01}
There is a $u$-$v$ path in the logical graph $G$ with bandwidth $d$
if and only if there is a physical path connecting $p(u)$ and $p(v)$
in the physical network, with bandwidth $d$, in the entire time
interval $[\tau, \tau+\delta]$.
\hfill$\Box$
\end{theorem}

We should note that the logical graph is dependent on the physical network,
the set of connection requests $\cR$, and the time interval
$[\tau, \tau+\delta]$.
%%%%%%%%%%%%%%%%%%%%%%%%%%%%%%%%%%%%%%%%%%%%%%%%
%%%%%%%%%%%%%%%%%%%%%%%%%%%%%%%%%%%%%%%%%%%%%%%%

%==============================================================================
\section{Algorithms for $\tOED$}
\label{sec:04}
%==============================================================================
%
\noindent
In this section, we present our novel solutions to $\tOED$ making use of the
logical graph introduced in Section~\ref{sec:03}.
In Section~\ref{sec:04A}, we present a polynomial time greedy algorithm.
In Section~\ref{sec:04B}, we present an integer linear programming (ILP)-based
approach for computing an optimal solution to $\tOED$.
In Section~\ref{sec:04C}, we discuss a variant of the ILP that can be used
to compute optimal entanglement distribution without using inter-satellite
links.

%==============================================================================
\subsection{Polynomial Time Greedy Algorithm}
\label{sec:04A}
%==============================================================================
%
\noindent
Our greedy algorithm is presented in Algorithm~\ref{alg1}.
It considers the requests in non-increasing order of the reward-demand ratio.
Whenever a request can be served, network resources are reserved for the
corresponding connection request.
%
%%
%==============================================================================
\begin{algorithm}[htbp]
\caption{{\rm \bf OED-Greedy}}
\label{alg1}
\LinesNumbered
%
%------------------------------------------------------------------------------
\KwIn{Network state information and a set of connection requests $\cR$}
\KwOut{A feasible subset $\cR_{greedy}$ of $\cR$ and the service path
$\pi_i$ for each $r_i \in \cR_{greedy}$}
%------------------------------------------------------------------------------
%
%
%------------------------------------------------------------------------------
Sort the requests in $\cR$ in non-decreasing order of
$\frac{w_i}{d_i}$.
WLOG, assume that
$\frac{w_1}{d_1} \ge \frac{w_2}{d_2} \ge \cdots \ge \frac{w_N}{d_N}$.
\label{alg1:L01}
%------------------------------------------------------------------------------

%------------------------------------------------------------------------------
Construct the logical graph $G(V, E)$ from the physical network and the set of
connection requests $\cR$.
\label{alg1:L02}
%------------------------------------------------------------------------------

%------------------------------------------------------------------------------
$\cRgreedy \leftarrow \emptyset$.
\label{alg1:L03}
%------------------------------------------------------------------------------

%------------------------------------------------------------------------------
\For{$i := 1$ \KwTo $N$}{
\label{alg1:L04}
%
%------------------------------------------------------------------------------
Compute a $l(s_i)$--$l(t_i)$ path $\pi_i$ of bandwidth at least $d_i$ in $G$.
Set $\pi_i \leftarrow \tNULL$ if such a path does not exist.
\label{alg1:L05}
%------------------------------------------------------------------------------

%------------------------------------------------------------------------------
\If{$\pi_i \ne \tNULL$}{
\label{alg1:L06}
%
%------------------------------------------------------------------------------
Add $r_i$ to $\cRgreedy$.
\label{alg1:L07}
%------------------------------------------------------------------------------

%------------------------------------------------------------------------------
Reserve the resources needed for path $\pi_i$.
\label{alg1:L08}
%------------------------------------------------------------------------------

%------------------------------------------------------------------------------
Update graph $G$ to its residual graph by removing the resources needed for
path $\pi_i$.
\label{alg1:L09}
%------------------------------------------------------------------------------
}
%------------------------------------------------------------------------------
}
%------------------------------------------------------------------------------

%------------------------------------------------------------------------------
Output $\cRgreedy$ and $\{\pi_i | r_i \in \cRgreedy\}$.
\label{alg1:L10}
%------------------------------------------------------------------------------
%
\end{algorithm}
%==============================================================================

Algorithm~\ref{alg1} has a worst-case time complexity bounded by a lower-order
polynomial.
After sorting, the main steps of the algorithm consist of $N$ path finding
processes, each of which can be accomplished via breadth-first-search.
While this algorithm does not guarantee to find an optimal solution,
it is very fast, and can find close to optimal solutions in most cases,
as demonstrated in our evaluations.

%==============================================================================
\subsection{ILP-based Optimal Algorithm}
\label{sec:04B}
%==============================================================================
%
\noindent
In order to evaluate the performance of Algorithm~\ref{alg1}, we design an
Integer Linear Programming (ILP)-based algorithm.
The main idea of our ILP-based algorithm is based on integer multi-commodity
flows.
For each request $r_i \in \cR$, we associate a binary variable $x_i$.
Here $x_i$ is $1$ if $r_i$ is served, and $0$ otherwise.
For each edge $(u, v) \in E$, we associate $N$ pairs of binary variables
$f_i(u, v)$, $f_i(v, u)$, $i = 1, 2, \ldots, N$.
Here $f_i(u, v)$ is the amount of type-$i$ flow from $u$ to $v$.
The ILP maximizes its linear objective function $\sum_{i=1}^{N} w_i \times x_i$,
subject to linear constraints described as follows.
%
%
%------------------------------------------------------------------------------
\emph{Flow out of a source node}:
For each $i \in \{1, 2, \ldots, N\}$, the net flow of type-$i$ out of vertex
$l(s_i)$ is $x_i$.
%
%
%------------------------------------------------------------------------------
\emph{Flow into a destination node}:
For each $i \in \{1, 2, \ldots, N\}$, the net flow of type-$i$ into vertex
$l(t_i)$ is $x_i$.
%
%
%------------------------------------------------------------------------------
\emph{Flow conservation}:
For each $i \in \{1, 2, \ldots, N\}$, for each vertex $z \in V$ that does
not correspond to $s_i$ or $t_i$,
i.e.,
$z \ne l(s_i)$
and
$z \ne l(t_i)$,
the net flow of type-$i$ into vertex $z$ is $0$.
%
%
%------------------------------------------------------------------------------
\emph{Channel capacity}:
For each edge $(u, v) \in E$,
$\sum_{i=1}^{N} (f_i(u, v) \times d_i + f_i(v, u) \times d_i)$ does not
exceed the number of available channels between nodes $p(u)$ and $p(v)$.
%
%
%------------------------------------------------------------------------------
\emph{Transmitters/Receivers/QMs}:
If $f_i(u, v) = 1$, we need to reserve $d_i$ transmitters at node
$p(u)$, $d_i$ receivers at node $p(v)$, and $d_i$ QMs
at both $p(u)$ and $p(v)$.
%
%
%==============================================================================

After solving the ILP described above, we can obtain an optimal solution
to the $\tOED$ problem.
The optimal set of requests to be served is given by
$\cR_{opt} = \{r_i \in \cR | x_i = 1\}$.
The path $\pi_i$ for $r_i \in \cR_{opt}$ can be obtained by tracing out the
edges $(u, v)$ where $f_i(u, v)$ is $1$.
%------------------------------------------------------------------------------

%==============================================================================
\subsection{Restricted ILP for Systems without Inter-satellite Links}
\label{sec:04C}
%==============================================================================
%
\noindent
To study the impact of allowing inter-satellite links in path computation, we also design
a restricted ILP-based algorithm.
The restricted ILP is a slight modification of the ILP studied in
Section~\ref{sec:04B}.
The only modification is not allowing inter-satellite links.
%%%%%%%%%%%%%%%%%%%%%%%%%%%%%%%%%%%%%%%%%%%%%%%%%%%%
%%%%%%%%%%%%%%%%%%%%%%%%%%%%%%%%%%%%%%%%%%%%%%%%%%%%

%==============================================================================
\section{Performance Evaluation}
\label{sec:05}
%==============================================================================
\noindent
In this section, we present the evaluation results of our space-ground
integrated quantum network model as well as our proposed algorithms.
The evaluation was done on a workstation running Ubuntu 22.04 system with
i9-12900 CPU and 64GB memory.
ILP instances were solved using the Gurobi optimizer.

\subsection{Evaluation Settings}
\label{sec:05A}
%==============================================================================
\noindent
{\bf Physical Network}:
%==============================================================================
We use $60$ ground stations, each located in a major city in the world.
Each physical network used in our evaluation consists of these $60$ ground
stations and $R \times K$ satellites, with $R=K$ taking a value in the set
$\{1, 2, \ldots, 25\}$.
We set $R$ and $K$ to be the same so that the satellites are evenly
distributed around the Earth.
We set the altitude of the satellites as $550$km and the orbital period as
$1.5$ hours, both of which are consistent with LEO satellites such
as Micius from the QUESS (Quantum Experiments at Space Scale) project
\cite{SatQIN2023, Sidhu2021Advances}. 
In our evaluation, we use the line-of-sight to determine the communication
range.
As a result, the inter-satellite communication range is $4988.11$km,
and the communication range between a satellite and a ground station
is $2703.81$km.
We assume the number of transmitters, receivers, and QMs at a node to be $10$. 
The number of available channels between two nodes within their
communication range is an integer between $1$ and $5$.

%==============================================================================
\noindent
{\bf Connection Requests}:
%==============================================================================
The source and destination nodes comprising each connection request are
chosen from the pool of $60$ ground stations.
The demand and reward for each connection request are integers between $1$ and $5$.
We used $N = 1, 10, 20$, and $30$ in our evaluations.
Start time $\tau$ takes a value in the set $\{0.0, 0.5, 1.0, 1.5, \ldots, 23.5\}$.
We will describe the choice of $\delta$ in Section~\ref{sec:05B}.

%==============================================================================
\subsection{Evaluation Scenarios}
\label{sec:05B}
\noindent
%==============================================================================
We evaluate our model and algorithms in the following three scenarios:
\begin{enumerate}
\item[(i)]
In this scenario, $\cR$ consists of one connection request, between
NYC and Singapore.
The demand is $1$ and the reward is $1$.
The start time is midnight.
We study two cases.
In the first case, we hold $R=K$ at $10$ and $20$, respectively,
and let $\delta$ vary from $0.00$ to $0.30$ in increments of $0.01$.
In the second case, we hold $\delta$ at $0$ and $0.1$, respectively,
and let $R=K$ vary from $1$ to $25$ in increments of $1$.
\textcolor{black}{This scenario is designed to study the impact of allowing
inter-satellite links on connecting two remote ground stations.
In the first case, we study the impact of $\delta$ (for fixed $R$ and $K$).
In the second case, we study the impact of $R$ and $K$ (for fixed $\delta$).}

\item[(ii)]
In this scenario, $\cR$ consists of $20$ connection requests, with the
source and destination randomly chosen from the pool of $60$ ground stations.
The demand and reward for each request are random integers between $1$ and $5$.
The start time is midnight for all $20$ connection requests.
We study two cases.
In the first case, we hold $R=K$ at $10$ and $20$, respectively, and
allow $\delta$ to vary from $0.00$ to $0.30$ in increments of $0.01$.
In the second case, we hold $\delta$ at $0.01$ and $0.1$, respectively, and
let $R=K$ vary from $1$ to $20$ in increments of $1$.
\textcolor{black}{This scenario is designed to study the impact of $\delta$
(for fixed $R$ and $K$)
and the impact of $R$ and $K$ (for fixed $\delta$),
in terms of total reward.}

\item[(iii)]
In this scenario, for each value of $N=10, 20$, and $30$, we generate
$48$ different sets of connection requests:
$\cR^{N}_{1}, \cR^{N}_{2}, \ldots, \cR^{N}_{48}$.
Each $\cR^{N}_{j}$ consists of $N$
connection requests where the source and destination are randomly
chosen from the $60$ ground stations,
and the demand and reward are random integers between 1 and 5.
All connection requests in $\cR^{N}_{j}$ have start time
$0.5 \times (j-1)$.
For each of $R=K= 10, 15$, and $20$, we evaluate the algorithms for each of $\delta=0.1, 0.05, 0.01$, and $0.001$.
This scenario is designed to conduct a more extensive performance evaluation 
of the proposed algorithms,
with the start times evenly distributed throughout the day.

\end{enumerate}

%==============================================================================
\subsection{Evaluation Results}
\label{sec:05C}
\noindent
%==============================================================================
In this section, we present evaluation results, together with our observations
and analysis.
We use $\tGreedy$, $\tILP$, and $\trILP$ to denote the
greedy algorithm, the ILP-based algorithm, and the restricted ILP-based
algorithm, respectively.

%==============================================================================
\begin{figure}[htbp]
\centering
\includegraphics[width=3.4in]{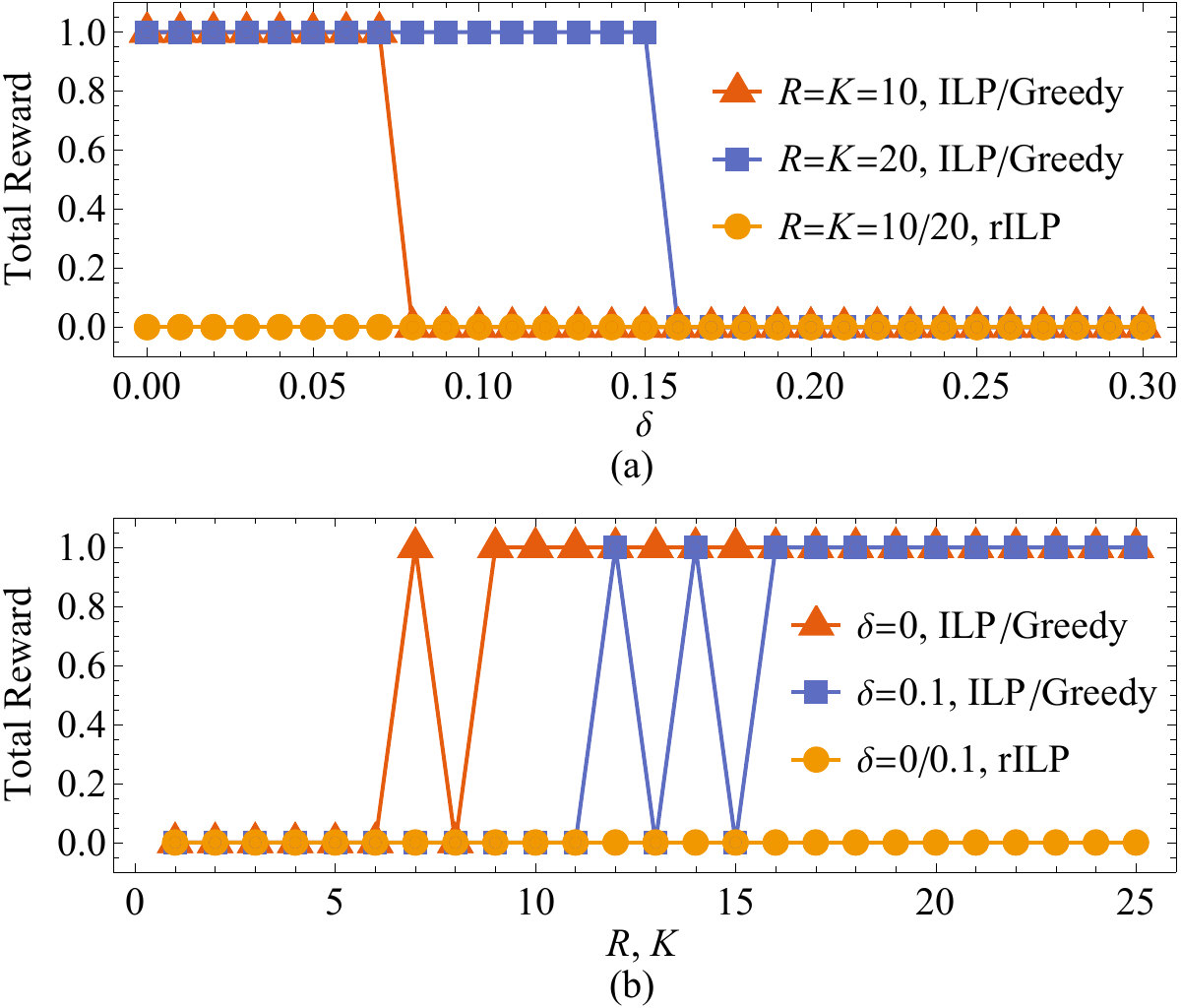}
\caption{Evaluation results of $\tGreedy$, $\tILP$, and $\trILP$ with varying $R$, $K$, and $\delta$ for a single NYC-Singapore request with demand $1$ and reward $1$.}
\label{fig:nyc-singapore}
\end{figure}
%==============================================================================
%==============================================================================
\noindent
{\bf Scenario (i):}
%==============================================================================
Fig.~\ref{fig:nyc-singapore} illustrates the results of Scenario (i).
Fig.~\ref{fig:nyc-singapore}(a) shows that $\tGreedy$ (and $\tILP$)
can find an NYC-Singapore path for
$\delta \le 0.07$ when $R=K=10$, and for
$\delta \le 0.15$ when $R=K=20$.
In contrast, $\trILP$ cannot find such a path for any $\delta=0.00, 0.01, \ldots, 0.30$ when both $R=K=10$ and $R=K=20$.
This is because no single satellite can have a communication link
with NYC and a communication link with Singapore at the same time.

Fig.~\ref{fig:nyc-singapore}(b) shows the result when we vary the value
of $R=K$, with $\delta$ fixed.
For $\delta=0$, when $R=K \le 6$, none of the algorithms can find an NYC-Singapore path.
When $R=K\ge 9$,
both $\tGreedy$ and $\tILP$ can find an NYC-Singapore path.
For $\delta=0.1$, when $R=K\le 11$, none of the
algorithms can find an NYC-Singapore path.
When $R=K\ge 16$,
both $\tGreedy$ and $\tILP$ can find an NYC-Singapore path.

Intuitively, when $R$ and $K$ increase, the number of edges in the
logical graph also increases.
But this is not always true.
Hence we see the fluctuations in the figure.
For $\delta=0$, $\tILP$ has a reward
value of $1$ with $R=K=7$, but the reward value drops to $0$ when $R=K=8$.
An interesting observation is that, for a fixed time interval
$[\tau, \tau+\delta]$,
the total reward corresponding to $\tILP$ is non-decreasing
when $R$ (and $K$) is increased to $n \times R$ (and $m \times K$)
for any integers $n \ge 1$ and $m \ge 1$.
This is because the resulting logical graph is a super-graph of
the original logical graph when $R$ (and $K$) is increased to
$n \times R$ (and $m \times K$).
For $\delta=0$, $\tILP$ has a reward value of
$1$ with $R=K=7$.
When $R$ and $K$ are increased to $14$ (a multiple of $7$), the
reward does not decrease.

\begin{figure}[htbp]
    \centering
    \includegraphics[width=3.4in]{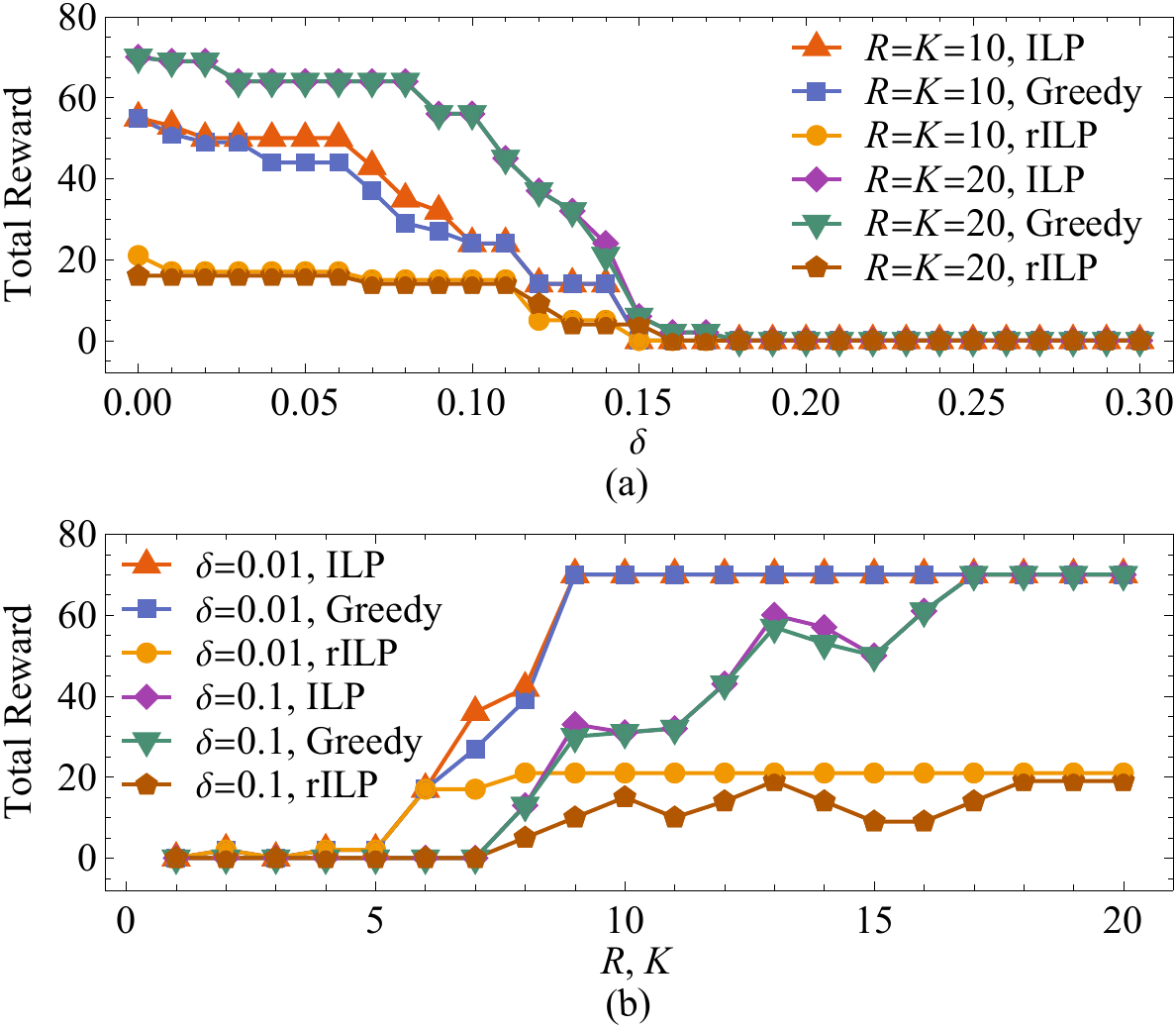}
    \caption{Evaluations results of $\tGreedy$, $\tILP$, and $\trILP$ with varying $R$, $K$, and $\delta$ for $20$ randomly generated requests.}
    \label{fig:20req}
\end{figure}
%==============================================================================
\noindent
{\bf Scenario (ii):}
%==============================================================================
Fig.~\ref{fig:20req} illustrates the results of Scenario (ii).
Fig.~\ref{fig:20req}(a) shows the impact of $\delta$ on the total reward
for each algorithm, with everything else fixed.
We observe that the total reward is non-decreasing as $\delta$ increases.
This is because fewer links will remain operational in the time interval
$[\tau, \tau+\delta]$ when $\delta$ increases.

Fig.~\ref{fig:20req}(b) shows the impact of $R$ (and $K$) on the total
reward for each algorithm, with $\delta$ fixed.
We observe that the total reward is non-decreasing (with some fluctuations) as
$\delta$ increases.
As explained in Scenario (i), the total reward will be non-decreasing
when $R$ (and $K$) is increased to a multiple of $R$ (and $K$).
Therefore, fluctuations are possible.

%==============================================================================
\noindent
{\bf Scenario (iii):}
%==============================================================================
Table~\ref{tab:table1} presents more extensive evaluation results.
Unlike Scenarios (i) and (ii), we have connection requests
with start time $\tau$ taking the values
$0.5 \times (j-1)$ for $j=1, 2, \ldots, 48$.
This means we have starting times at every hour and every half hour.

Since the test cases have different time intervals, it is less meaningful
to compute the total reward across all test cases.
Rather, for each test case, we compute the ratio of the reward for
$\tGreedy$ over that for $\tILP$, as well as the ratio of the reward for
$\trILP$ over that for $\tILP$, and take the average of these
ratios.
We also record the average running times for $\tGreedy$ and $\tILP$ (the running time for $\trILP$ is similar to that for $\tILP$).
We observe that $\tGreedy$ requires much less time than $\tILP$,
while computing solutions nearly as good as those computed by $\tILP$.

%=============================================================================== 

%=============================================================================== 
\begin{table}[htbp]
\scriptsize
\centering
\caption{Performance of $\tGreedy$, $\tILP$, and $\trILP$ for $1728$ test cases.}
\label{tab:table1}
\begin{tabular}{l r r | r | r r | r}
\toprule
\multirow{2}{*}{$N$} & \multirow{2}{*}{$R,K$} & \multirow{2}{*}{$\delta$} & \textbf{rILP} & \multicolumn{2}{c|}{\textbf{Greedy}} & \textbf{ILP}\\

& & & \textbf{Ratio}  & \textbf{Ratio} & \textbf{Time} &  \textbf{Time} \\\hline

\multirow{12}{*}{\textbf{10}} & \multirow{4}{*}{$10$} & $0.1$ & $0.1199$ & $0.9966$ & $0.0005s$ & $0.1462s$ \\\cline{3-7}

&  & $0.05$ & $0.1359$ & $0.9831$ & $0.0006s$ & $0.1600s$ \\\cline{3-7}

&  & $0.01$ & $0.1749$ & $0.9982$ & $0.0006s$ & $0.1733s$ \\\cline{3-7}

&  & $0.001$ & $0.1809$ & $0.9990$ & $0.0006s$ & $0.1789s$ \\\cline{2-7}

& \multirow{4}{*}{$15$} & $0.1$ & $0.0915$ & $0.9860$ & $0.0018s$ & $0.6498s$ \\\cline{3-7}

&  & $0.05$ & $0.1803$ & $1.0000$ & $0.0021s$ & $0.7640s$ \\\cline{3-7}

&  & $0.01$ & $0.2210$ & $1.0000$ & $0.0024s$ & $0.8754s$ \\\cline{3-7}

&  & $0.001$ & $0.2272$ & $1.0000$ & $0.0024s$ & $0.8875s$ \\\cline{2-7}

&  \multirow{4}{*}{$20$} & $0.1$ & $0.1059$ & $1.0000$ & $0.0077s$ & $3.5907s$ \\\cline{3-7}

&  & $0.05$ & $0.1945$ & $1.0000$ & $0.0089s$ & $4.2291s$ \\\cline{3-7}

&  & $0.01$ & $0.2222$ & $1.0000$ & $0.0098s$ & $4.7088s$ \\\cline{3-7}

&  & $0.001$ & $0.2287$ & $1.0000$ & $0.0101s$ & $4.9035s$ \\\cline{1-7}

\multirow{12}{*}{\textbf{20}} & \multirow{4}{*}{$10$} & $0.1$ & $0.1393$ & $0.9756$ & $0.0009s$ & $0.2715s$ \\\cline{3-7}

&  & $0.05$ & $0.1329$ & $0.9643$ & $0.0011s$ & $0.3185s$ \\\cline{3-7}

&  & $0.01$ & $0.1606$ & $0.9932$ & $0.0012s$ & $0.3430s$ \\\cline{3-7}

&  & $0.001$ & $0.1700$ & $0.9936$ & $0.0012s$ & $0.3519s$ \\\cline{2-7}

& \multirow{4}{*}{$15$} & $0.1$ & $0.0868$ & $0.9683$ & $0.0034s$ & $1.3464s$ \\\cline{3-7}

&  & $0.05$ & $0.1576$ & $1.0000$ & $0.0038s$ & $1.5269s$ \\\cline{3-7}

&  & $0.01$ & $0.1980$ & $1.0000$ & $0.0043s$ & $1.7278s$ \\\cline{3-7}

&  & $0.001$ & $0.2056$ & $1.0000$ & $0.0044s$ & $1.7764s$ \\\cline{2-7}

&  \multirow{4}{*}{$20$} & $0.1$ & $0.0959$ & $1.0000$ & $0.0146s$ & $6.9379s$ \\\cline{3-7}

&  & $0.05$ & $0.1719$ & $1.0000$ & $0.0168s$ & $8.1285s$ \\\cline{3-7}

&  & $0.01$ & $0.2074$ & $1.0000$ & $0.0186s$ & $9.0636s$ \\\cline{3-7}

&  & $0.001$ & $0.2128$ & $1.0000$ & $0.0191s$ & $9.2650s$ \\\hline

\multirow{12}{*}{\textbf{30}} & \multirow{4}{*}{$10$} & $0.1$ & $0.1618$ & $0.9523$ & $0.0014s$ & $0.4592s$ \\\cline{3-7}

&  & $0.05$ & $0.1296$ & $0.9311$ & $0.0017s$ & $0.5930s$ \\\cline{3-7}

&  & $0.01$ & $0.1517$ & $0.9685$ & $0.0018s$ & $0.6046s$ \\\cline{3-7}

&  & $0.001$ & $0.1569$ & $0.9736$ & $0.0018s$ & $0.6210s$ \\\cline{2-7}

& \multirow{4}{*}{$15$} & $0.1$ & $0.0834$ & $0.9455$ & $0.0055s$ & $2.3433s$ \\\cline{3-7}

&  & $0.05$ & $0.1441$ & $0.9978$ & $0.0074s$ & $2.9550s$ \\\cline{3-7}

&  & $0.01$ & $0.1856$ & $1.0000$ & $0.0084s$ & $3.4108s$ \\\cline{3-7}

&  & $0.001$ & $0.1945$ & $1.0000$ & $0.0086s$ & $3.4192s$ \\\cline{2-7}

&  \multirow{4}{*}{$20$} & $0.1$ & $0.0965$ & $0.9981$ & $0.0215s$ & $10.6800s$ \\\cline{3-7}

&  & $0.05$ & $0.1649$ & $1.0000$ & $0.0247s$ & $14.7769s$ \\\cline{3-7}

&  & $0.01$ & $0.1929$ & $1.0000$ & $0.0274s$ & $16.4577s$ \\\cline{3-7}

&  & $0.001$ & $0.1992$ & $1.0000$ & $0.0281s$ & $16.9239s$ \\

\bottomrule
\end{tabular}
\end{table}
%%%%%%%%%%%%%%%%%%%%%%%%%%%%%%%%%%%%%%%%%%%%%%%%%%%%%%
%%%%%%%%%%%%%%%%%%%%%%%%%%%%%%%%%%%%%%%%%%%%%%%%%%%%%%

%==============================================================================
\section{Conclusions}
\label{sec:06}
%==============================================================================
\noindent
In this paper, we have proposed a system model for a space-ground integrated
quantum network.
Unlike previous research, we take into consideration the movement of
the satellites and inter-satellite links.
We also ensure that all links used in the computed entanglement paths
are guaranteed to be operational within an entire time interval.

To study entanglement distribution in such a dynamic network, we
propose a novel concept of logical graphs.
We design a polynomial time greedy algorithm for solving the entanglement
distribution problem in a dynamic physical network with the aid of the
corresponding logical graph.

We demonstrate that it is possible to establish an NYC-Singapore path using inter-satellite links.
Extensive evaluations show that our greedy algorithm can compute solutions
that are nearly as good as optimal, while using much less time.
%%%%%%%%%%%%%%%%%%%%%%%%%%%%%%%%%%%%%%%%%%%%%%%%%%%%%
%%%%%%%%%%%%%%%%%%%%%%%%%%%%%%%%%%%%%%%%%%%%%%%%%%%%%

%%%%%%%%%%%%%%%%%%%%%%%%%%%%%%%%%%%%%%%%%%%%%%%%%%%%%%
%%%%%%%%%%%%%%%%%%%%%%%%%%%%%%%%%%%%%%%%%%%%%%%%%%%%%%

\end{document}